\documentclass{amsart}
\usepackage{amssymb,latexsym}
\usepackage{graphicx}
\numberwithin{equation}{section}

\begin{document}
 \hfill{\today}

\title{The Airy transform and the associated polynomials}
\author{D. Babusci$^\dag$, G. Dattoli$^\ddag$, D. Sacchetti$^\diamond$} 

\address{$^\dag$ INFN - Laboratori Nazionali di Frascati, via E. Fermi 40, I-00044 Frascati.}
\email{danilo.babusci@lnf.infn.it}

\address{$^\ddag$ ENEA - Dipartimento Tecnologie Fisiche e Nuovi Materiali, Centro Ricerche Frascati\\
                 C. P. 65, I-00044 Frascati.}
 \email{giuseppe.dattoli@enea.it}
                 
\address{$^\diamond$ ENEA - Dipartimento di Statistica, Probabilit\`a e Statistica Applicata, Universit\`a 
                  "Sapienza" di Roma, P.le A. Moro, 5, 00185 Roma.}
\email{dario.sacchetti@uniroma1.it}

\begin{abstract}
The Airy transform is an ideally suited tool to treat problem in classical and quantum optics. Even though 
the relevant mathematical aspects have been thoroughly investigated, the possibility it offers are wide and 
some aspects, as the link with special functions and polynomials, still contains unexplored aspects. In this 
note we will show that the so called Airy polynomials are essentially the third order Hermite polynomials. 
We will also prove that this identification opens the possibility of developing new conjectures on the 
properties  of this family of polynomials.
\end{abstract}

\maketitle

\section{Introduction}
The theory of ordinary and generalized Hermite polynomials has largely benefited of the operational 
formalism. The two variable Hermite-Kamp\'e de F\'eri\'et polynomials \cite{Appe} can be defined using 
the following identity \cite{Dat1}:
\begin{equation}
\label{eq:herm}
H_n (x,y) \,=\, \mathrm{e}^{y \partial_x^2}\,x^n
\end{equation}
which involves the action of an exponential operator, containing a second order derivative, on a monomial. 
The explicit form of the polynomials $H_n (x,y)$ can be obtained by means of a straightforward expansion 
of the exponential in eq. \eqref{eq:herm}, which yields:
\begin{equation}
H_n (x,y) \,=\, \sum_{r = 0}^\infty\, \frac{y^r}{r!}\,\partial_x^{2r} x^n \,=\, n!\,\sum_{r = 0}^{\left[n/2\right]} \,
\frac{x^{n - 2r}\,y^r}{(n - 2r)!\,r!}\;,
\end{equation}
where the variables $x, y$ independent each other\footnote{By interpreting the variable $y$ as a 
parameter, the standard Hermite form are recovered by the identities 
$H_n(2x, -1) = H_n (x)$ and $H_n (x, -1/2) = He_n (x)$. }. 
By keeping, therefore, the derivative of both sides of eq. \eqref{eq:herm} with respect to  $y$, we find 
that the Hermite polynomials can be viewed as the solution of the following heat equation
\begin{equation}
\partial_y F(x,y) \,=\, \partial_x^2 F(x,y)\;, \qquad \qquad F(x,0) \,=\, x^n\;.
\end{equation}
For $y >0$ they can be written in terms of the Gauss-Weierstrass transform \cite{Wolf}:
\begin{equation}
\label{eq:gwtr}
H_n (x,y) \,=\, \frac{1}{2\sqrt{\pi y}}\,\int_{- \infty}^\infty\, \mathrm{d}\xi\,\xi^n\,
\exp{\left\{-\frac{(x - \xi)^2}{4 y}\right\}}\;.
\end{equation}
which is a standard mean of solutions for the heat type problems.

The higher order Hermite polynomials \cite{Dat1}, widely exploited in combinatorial quantum field 
theory \cite{Horz}, can be expressed as a generalization of the operational identity \eqref{eq:herm}, 
and indeed they write\footnote{The upper index $m$, denoting the order of the polynomials, is omitted 
for Hermite polynomials of order 2.}
\begin{equation}
\label{eq:lach}
H_n^{(m)} (x,y) \,=\, \mathrm{e}^{y \partial_x^m}\,x^n \,=\, n!\,\sum_{r = 0}^{\left[n/m\right]}  \,
\frac{x^{n - mr}\,y^r}{(n - mr)!\,r!}\;.
\end{equation}
Therefore, we can ask whether an integral transform, a sort of  generalization of the Gauss-Weierstrass 
transform,  also holds for the higher order case. We start discussing the case of Hermite polynomials of even 
order with negative values of the $y$ parameter, namely:
\begin{equation}
\label{eq:hneg}
H_n^{(2p)} (x,-|y|) \,=\, \mathrm{e}^{-|y|  \partial_x^{2p}}\,x^n\;,
\end{equation}
We express this family of polynomials in terms of a suitable transform following the procedure, put forward 
in \cite{Dat2}, which considers the operator function
\begin{equation}
\hat{F} \,=\, f(\partial_x)\;,
\end{equation}
where $f(x)$ is a function admitting a Fourier transform $\tilde{f}(k)$. With this assumption we find that the 
operator $\hat{F}$ can be written as: 
\begin{equation}
\hat{F} \,=\, \frac{1}{\sqrt{2 \pi}}\,\int_{- \infty}^{\infty}\, \mathrm{d}k\, \tilde{f}(k) \mathrm{e}^{\imath k\partial_x}\;,
\end{equation}
and, therefore, we can express the action of the operator $\hat{F}$ on a given function $g(x)$ as the integral 
transform indicated below
\begin{equation}
\hat{F} g(x) \,=\, \frac{1}{\sqrt{2 \pi}}\,\int_{- \infty}^{\infty}\, \mathrm{d}k\, \tilde{f}(k)\,
\mathrm{e}^{\imath k\partial_x} g(x) \,=\,\frac{1}{\sqrt{2 \pi}}\,\int_{- \infty}^{\infty}\, \mathrm{d}k\, 
\tilde{f}(k)\,g(x + \imath k)\;.
\end{equation}
We can now apply the same procedure to express the exponential operator intervening in the definition of 
$H_n^{(m)}$ (see eq. \eqref{eq:lach}), thus obtaining the following integral transform yielding the even order 
Hermite polynomials
\begin{equation}
H_n^{(2p)} (x,-|y|) \,=\, \frac{1}{\sqrt{2 \pi}}\,\int_{- \infty}^{\infty}\, \mathrm{d}k\, \tilde{e}_{2p} (k)\, 
\left(x - \imath k \sqrt[2p]{|y|}\right)^n
\end{equation}
with
\begin{equation}
\tilde{e}_{2p} (k) \,=\, \frac{1}{\sqrt{2 \pi}}\,\int_{- \infty}^{\infty}\, \mathrm{d}x\, \mathrm{e}^{- x^{2p}}\, 
\mathrm{e}^{- \imath k x}\;.
\nonumber
\end{equation}
which\footnote{The choice of discussing even order $m$ is motivated by the the request that the integrals defining 
$\tilde{e}_{m} (k)$ is convergent.}, after a redefinition of the variable, can also be written as
\begin{equation}
H_n^{2p)} (x,-|y|) \,=\,- \frac{1}{\sqrt{2 \pi}}\,\frac{1}{\imath \sqrt[2p]{|y|}}\, \int_{- \infty}^{\infty}\, 
\mathrm{d}\xi\, \xi^n \,\tilde{e}_{2p}\, \left(\frac{x - \xi}{\imath \sqrt[2p]{|y|}}\right)\;.
\end{equation}
It must remarked that the same procedure, applied to the case $m = 2$, does not lead to a Gauss-Weierstrass 
transform as in eq. \eqref{eq:gwtr}, which holds only for $y > 0$. 

In the next section we will extend the formalism developed in these introductory remarks, and prove that the Airy 
transform and the associated polynomials can be framed within the same context.

\vspace{1cm}
\section{The Airy transform and the Airy polynomials}
The higher order Hermite polynomials satisfy the following recurrences \cite{Dat1}:
\begin{eqnarray}
\label{eq:recu}
\partial_x H_n^{(m)} (x,y) \!\!&=&\!\! n H_{n-1}^{(m)} (x,y) \nonumber \\
H_{n+1}^{(m)} (x,y) \!\!&=&\!\! x H_{n}^{(m)} (x,y) + m\, \frac{n!}{(n-m+1)!}\, H_{n-m+1}^{(m)} (x,y) \\
\partial_y H_n^{(m)} (x,y) \!\!&=&\!\! \frac{n!}{(n-m)!}\, H_{n-m}^{(m)} (x,y)\;. \nonumber
\end{eqnarray}
with the combination of the first and third recurrences yielding
\begin{equation}
\partial_y H_n^{(m)} (x,y) \,=\, \partial_x^m H_n^{(m)} (x,y)\;, \qquad\qquad H_{n}^{(m)} (x,0) \,=\, x^n\,, 
\end{equation}
Therefore, the higher order Hermite polynomials satisfy a generalized heat equation, and this justifies the 
operational definition given in eq. \eqref{eq:lach}. Furthermore, by interpreting $y$ as a parameter, we can 
use the first two recurrences to prove tha they satisfy the $m$-th order ODE:
\begin{equation}
\left(y \frac{\mathrm{d}^m}{\mathrm{d}x^m}\,+\,x  \frac{\mathrm{d}}{\mathrm{d}x}\right) H_n^{(m)} (x,y) 
\,=\, n H_n^{(m)} (x,y)
\end{equation}

In the previous section we have considered even order Hermite polynomials only. Here we will discuss the 
third order case and their important relationship with the Airy transform  and the Airy polynomials \cite{Vall}. 
Before getting into this specific aspect of the problem, we note that the following identity holds:
\begin{equation}
\label{eq:elam}
\mathrm{e}^{\lambda x^3} \,=\, \int_{- \infty}^{\infty}\, \mathrm{d}t\, \mathrm{e}^{\sqrt[3]{\lambda} xt}\, Ai(t)\;,
\end{equation}
where
\begin{equation}
\label{eq:airy}
Ai (t) \,=\, \frac{1}{2 \pi}\,\int_{- \infty}^{\infty}\, \mathrm{d}\xi\, 
\exp{\left\{\frac{\imath}{3} \xi^3 + \imath\,x\,\xi \right\}}\;,
\end{equation}
is the Airy function \cite{Widd}.

Let us now consider the third order PDE
\begin{equation}
\partial_y F(x,y) \,=\, \partial_x^3 F(x,y)\;, \qquad\qquad F(x,0) \,=\, f(x)\;,
\end{equation}
whose formal solution writes
\begin{equation}
F(x,y) \,=\, \mathrm{e}^{y \partial_x^3} f(x)\;.
\end{equation}
By applying the identity given in eq. \eqref{eq:elam}, and by limiting ourselves to the case $y >0$, we find:
\begin{eqnarray}
\label{eq:fexp}
F(x,y) \!\!&=&\!\! \frac{1}{\sqrt[3]{3}}\, \int_{- \infty}^{\infty}\, \mathrm{d}k\, Ai\left(\frac{k}{\sqrt[3]{3}}\right)\, 
                             \mathrm{e}^{\sqrt[3]{y} k \partial_x}\, f(x) \nonumber \\
            \!\!&=&\!\! \frac{1}{\sqrt[3]{3}}\, \int_{- \infty}^{\infty}\, \mathrm{d}k\, Ai\left(\frac{k}{\sqrt[3]{3}}\right)\,
                            f(x + \sqrt[3]{y} k) \nonumber \\
            \!\!&=&\!\!  \frac{1}{\sqrt[3]{3 y}}\, 
                             \int_{- \infty}^{\infty}\, \mathrm{d}\xi\, Ai\left(- \frac{x - \xi}{\sqrt[3]{3 y}}\right)\, f(\xi)\;,
                             \qquad \qquad (y > 0)
\end{eqnarray}
which is recognized as the Airy transform of the function $f(x)$. The concept of the Airy transform  was 
introduced in ref. \cite{Widd}, and has found noticeable applications in classical and quantum physics 
\cite{Vall}. The Airy transform of a monomial, namely
\begin{equation}
\label{eq:apol}
\alpha i_n (x, y) \,=\, \frac{1}{\sqrt[3]{3 y}}\, \int_{- \infty}^{\infty}\, \mathrm{d}\xi\, 
                                   Ai\left(- \frac{x - \xi}{\sqrt[3]{3 y}}\right)\, \xi^n\;, \qquad \qquad (y > 0)
\end{equation} 
has been defined as Airy polynomials \cite{Vall,Widd}, but, according to the discussion of the previous section, 
they are also recognized as the third order Hermite polynomials $H_n^{(3)} (x,y)$. The characteristic 
recurrences  \eqref{eq:recu} (specialized to the case $m = 3$), can also be directly inferred from 
eq. \eqref{eq:apol}.

\vspace{1cm}
\section{Concluding remarks}
\begin{figure}
\begin{center}
\vspace{-2.5cm}
\includegraphics[width=3.5in]{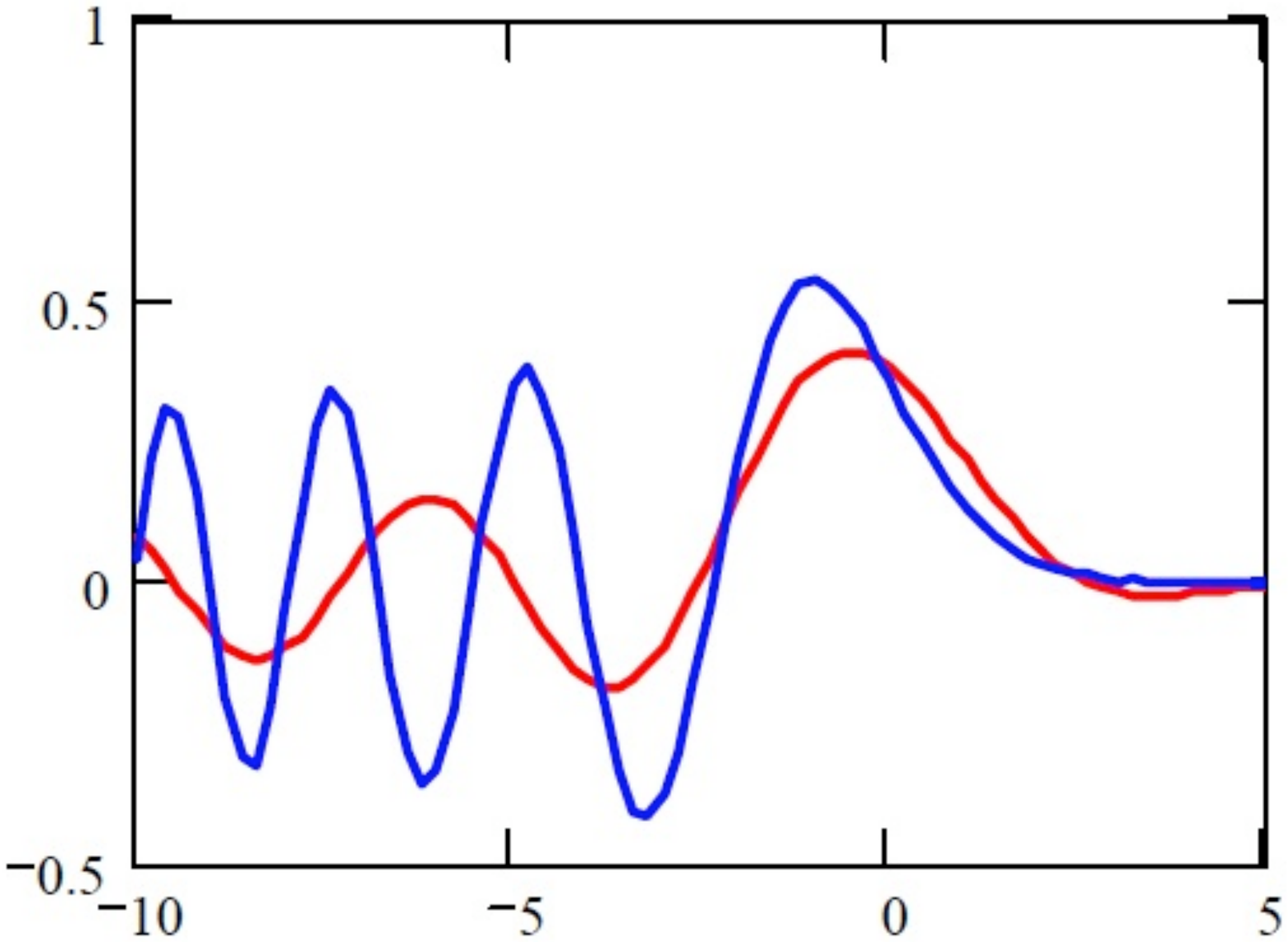}
\vspace{-2.5cm}
\caption{Comparison between the seventh order Airy function (red curve)
and the ordinary Airy function (blue curve).}
\label{fig:Aicmp}
\end{center}
\end{figure}

For further convenience we will introduce the following two variable extension of the Airy functions
\begin{equation}
\label{eq:gair}
Ai (x,y) \,=\, \frac{1}{2 \pi}\,\int_{- \infty}^{\infty}\, \mathrm{d}\xi\, 
\exp{\left\{\imath\,y\,t^3 + \imath\,x\,t \right\}}\;,
\end{equation}
that, in terms of the ordinary Airy function writes
\begin{equation}
Ai (x,y) \,=\, \frac{1}{\sqrt[3]{3y}}\,Ai\left(\frac{x}{\sqrt[3]{3y}}\right)\;.
\end{equation}
It is also easily shown that it satisfies the ODE
\begin{equation}
3\,y\,\frac{\mathrm{d}^2}{\mathrm{d}x^2} Ai(x,y) \,-\, x\,Ai(x,y) \,=\,0
\end{equation}
and that any other function linked to $Ai (x,y)$ by
\begin{equation}
Ai^{(m)}(x,y,z) \,=\, \mathrm{e}^{z \partial_x^m}\,Ai(x,y)
\end{equation}
satisfies the ODE
\begin{equation}
3\,y\,\frac{\mathrm{d}^2}{\mathrm{d}x^2} Ai^{(m)}(x,y,z) \,-\, \mathrm{e}^{z \partial_x^m}\,
\left[x\,Ai(x,y,z)\right] \,=\,0
\end{equation}
which on account of the identity
\begin{equation}
\mathrm{e}^{z \partial_x^m}\,x \,=\,  \left(x \,+\, m\,z\,\partial_x^{m-1}\right)\,
\mathrm{e}^{z \partial_x^m}\
\end{equation}
can be written as
\begin{equation}
\left(m\,z\,\frac{\mathrm{d}^{m - 1}}{\mathrm{d}x^{m - 1}}\,-\,3\,y\,\frac{\mathrm{d}^2}{\mathrm{d}x^2}
\,+\,x\right) Ai^{(m)} (x,y,z) \,=\, 0
\end{equation}
If we assume $m = 2p +1$, $y = 0$, and $mz = 1$,  we get the following generalization of the Airy 
function 
\begin{equation}
Ai^{(2p + 1)}(x) \,=\, \frac{1}{2 \pi}\,\int_{- \infty}^{\infty}\, \mathrm{d}t\, 
\exp{\left\{\frac{\imath}{2p + 1}\,t^{2p + 1} + \imath\,x\,t \right\}}\;,
\end{equation}
satisfying the ODE
\begin{equation}
\left(\frac{\mathrm{d}^{2p}}{\mathrm{d}x^{2p}}\,+\,x\right)\,Ai^{(2p+1)}(x) \,=\, 0\;.
\end{equation}
As an example a comparison between the ordinary Airy function and its generalization of order 7 is 
shown in Fig. \ref{fig:Aicmp}

The point of view developed in sec. 2 can be generalized to the case with $m > 3$.  We consider 
the generalized Hermite polynomials of odd order ($m = 2p + 1$) and note that the same procedure 
exploited in the previous sections leads to the following integral representation
\begin{equation}
H_n^{2p + 1)} (x,-|y|) \,=\,\frac{1}{\sqrt[2p + 1]{(2p + 1) |y|}}\, \int_{- \infty}^{\infty}\, \mathrm{d}\xi\, 
Ai^{(2p + 1)} \left(\frac{\xi - x}{\sqrt[2p + 1]{(2p + 1) |y|}}\right)\,\xi^n\;,
\end{equation}

The use of a further variable or parameter in the theory of special functions and/or polynomials 
offers a further degree of freedom, which may be helpful to derive new properties, otherwise 
hidden by the loose of symmetry deriving from the fact that a specific choice of the variable has 
been done. This is indeed the case of the Hermite polynomials, which acquire a complete new 
flavor by the use of the $y$ variable and the case of the Airy function given in eq. \eqref{eq:gair} 
which is easily shown to be the natural solution of the PDE
\begin{equation}
\label{eq:apde}
\partial_y Ai (x,y) \,=\, - \partial_x^3 Ai(x,y)\;,
\end{equation}
and the translation property
\begin{equation}
\mathrm{e}^{- z \partial_x^3} Ai(x,y) \,=\, Ai(x,y+z)\;.
\end{equation}
According to eq. \eqref{eq:apde}, we can write the solution of the equation
\begin{equation}
\partial_y F (x,y) \,=\, - \partial_x^3 F(x,y)\;, \qquad\qquad F(x,0) \,=\, g(x)\;,
\end{equation}
as follows
\begin{equation}
F(x,y) \,=\, \int_{- \infty}^{\infty}\, \mathrm{d}\xi\, Ai(x - \xi,y)\,g(\xi)\,.
\end{equation}

It is evident that the further generalization
\begin{equation}
Ai^{(2p + 1)} (x,y) \,=\, \frac{1}{2 \pi}\,\int_{- \infty}^{\infty}\, \mathrm{d}t\, 
\exp{\left\{\imath\,y\,t^{2p + 1}\,+\,\imath\,x\,t\right\}}\;, \qquad\qquad (y > 0)\;,
\end{equation}
satisfies a higher order heat equation and can be exploited to obtain the solution of the same 
family of equation in terms of an appropriate integral transform.

Before concluding this note let us consider the following Schr\"odinger equation 
\begin{equation}
\imath\,\hbar\,\Psi(x,t) \,=\, - \frac{\hbar^2}{2 m}\,\partial_x^2\,\Psi(x,t)\,+\,F\,x\,\Psi(x,t)\;,
\qquad\qquad \Psi(x,0) \,=\, \psi(x)\;,
\end{equation}
describing the motion of a particle in a linear potential ($F$ is a constant with the dimension 
of a force). This equation can be written in the more convenient form
\begin{equation}
\label{eq:scho}
\imath\,\partial_{\tau}\,\Psi(x,\tau) \,=\, - \partial_x^2\,\Psi(x,\tau) \,+\, b\,x\,\Psi(x,\tau)\;,
\qquad\qquad \Psi(x,0) \,=\, \psi(x)\;,
\end{equation}
where\footnote{This choice of the variables implies that $\tau$ has the dimensions of a squared 
length and $b$ of an inverse cube length, so that $b\tau^2$ has the dimensions of a length. }
\begin{equation}
\tau \,=\, \frac{\hbar t}{2 m}\;,\qquad\qquad b \,=\, \frac{2 F m}{\hbar^2}\;.
\end{equation}

As discussed in refs. \cite{Feng}, the previous equation can be treated using different means, 
a very simple solution being offered by the use of the following auxiliary function:
\begin{equation}
\Phi(x,\tau) \,=\, \mathrm{e}^{\frac{1}{3b} \partial_x^3}\,\Psi(x,\tau)\,
\end{equation}
which satisfies the equation
\begin{equation}
\imath\,\partial_{\tau}\,\Phi(x,\tau) \,=\, b\,x\,\Phi(x,\tau)\;.
\end{equation}
The solution of eq. \eqref{eq:scho} can therefore be written as
\begin{equation}
\Psi(x,t) \,=\, \hat{A}\,\psi(x)\;,\qquad 
\hat{A} \,=\, \mathrm{e}^{-\frac{1}{3b} \partial_x^3}\,\mathrm{e}^{- \imath b x \tau}\,
\mathrm{e}^{\frac{1}{3b} \partial_x^3}\;,
\end{equation}
where the operator $\hat{A}$, once written in an integral form, is recognized as the Airy transform.

\begin{figure}
\begin{center}
\vspace{-2.5cm}
\includegraphics[width=3.5in]{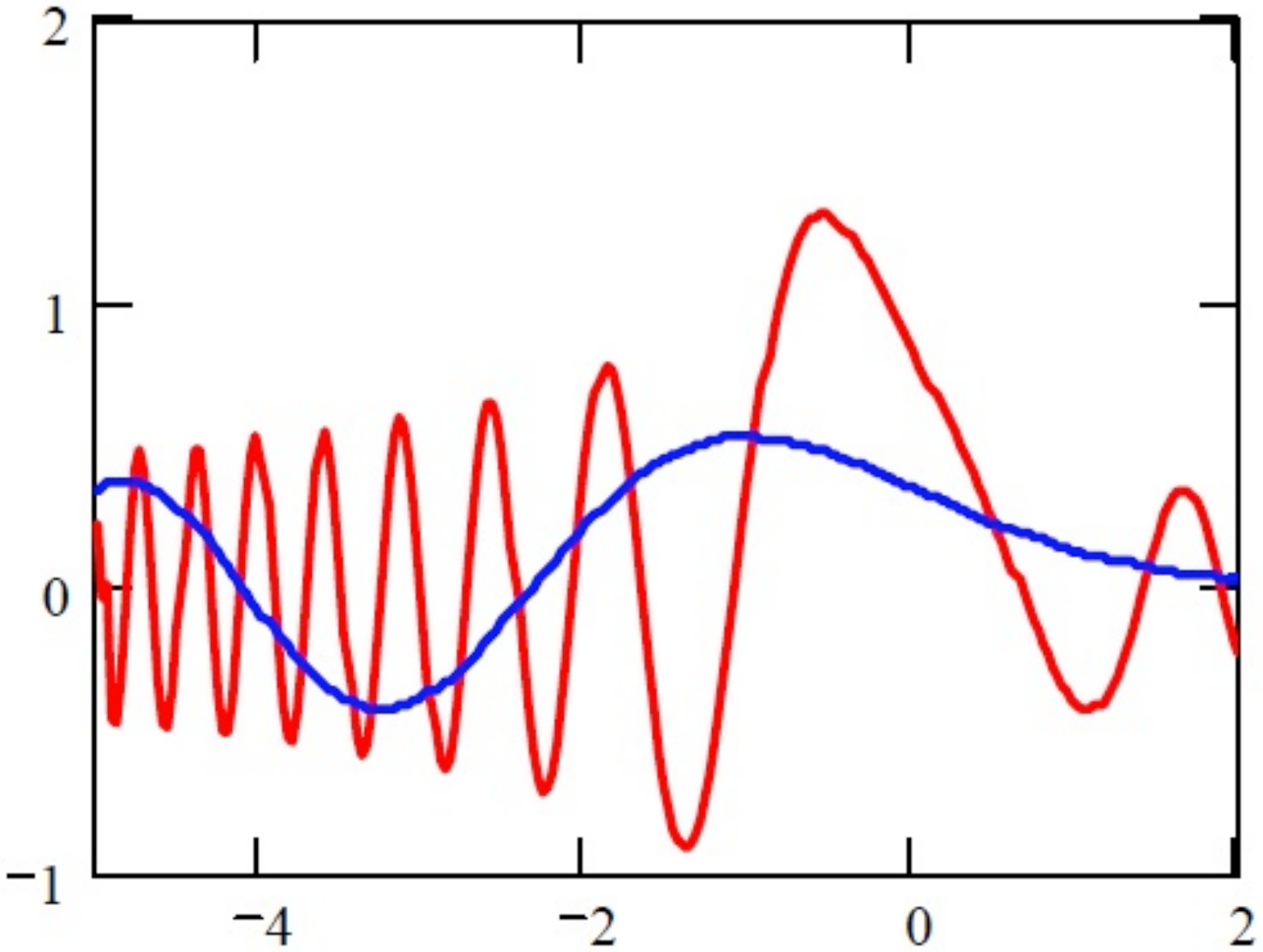}
\vspace{-2.5cm}
\caption{Comparison between the Watson Airy function $W(x)$ (red curve)
              and the ordinary Airy function (blue curve).
}
\label{fig:WAcmp}
\end{center}
\end{figure}

The solution of the equation 
\begin{equation}
\imath\,\partial_{\tau}\,\Psi(x,\tau) \,=\, - \partial_x^{2p}\,\Psi(x,\tau) \,+\, b\,x\,\Psi(x,\tau)\;,
\qquad\qquad \Psi(x,0) \,=\, \psi(x)\;,
\end{equation}
can be obtained in an analogous way, and it is given by
\begin{equation}
\Psi(x,t) \,=\, \hat{A}^{(2p)}\,\psi(x)\;,\qquad 
\hat{A}^{(2p)} \,=\, \mathrm{e}^{-\frac{1}{(2p + 1)b} \partial_x^{2p + 1}}\,\mathrm{e}^{- \imath b x \tau}\,
\mathrm{e}^{\frac{1}{(2p + 1) b} \partial_x^{2p + 1}}\;,
\end{equation}
where the operator $\hat{A}^{(2p)}$ can be expressed in terms of integral transforms, linked to the 
higher order Airy functions discussed in these sections.
Further examples of generalization of the Airy function are due to Watson \cite{Wats}, who introduced 
a complete class of functions with interesting properties for physical applications. As an example we 
consider the function expressed as
\begin{equation}
W(x) \,=\, \int_0^\infty\,\mathrm{d}t \,\cos(t^4\,+\,4 x t\,+\,2 x^2)
\end{equation}
which is compared to the ordinary Airy function in Fig. \ref{fig:WAcmp}, and satisfies the ODE
\begin{equation}
\frac{\mathrm{d}^2}{\mathrm{d}x^2}\,W(x) \,+\, 4\,x^2\,W(x) \,=\, 0\;.
\end{equation}
The use of the unitary transformation
\begin{equation}
\label{eq:Jtra}
J(x) \,=\, \mathrm{e}^{\frac{\imath}{4} \partial_x^2}\,W(x)\;,
\end{equation}
shows that it can be associated to the first ODE
\begin{equation}
x^2\,J(x)\,+\,\imath\left(x\,\frac{\mathrm{d}}{\mathrm{d}x}\,+\,\frac{1}{2}\right)\,J(x) \,=\, 0\;.
\end{equation}
The transformation \eqref{eq:Jtra} has removed the second order derivative and has simplified the 
underlying group of symmetry, which has been reduced from $SU(1,1)$ to the dilatation group. 
We have mentioned this example because it may open a new interesting point of view to Schr\"odinger 
type equations involving quadratic potentials.

Before concluding this paper we discuss whether the Airy polynomials can be exploited to obtain a series 
expansion of a given function. The problems associated with the orthogonal properties of ordinary and 
higher order Hermite polynomials has been thoroughly considered in a previous publications (see refs. 
\cite{Haim,Dat3}); here we will use the point of view developed in ref. \cite{Dat3}.
We consider indeed the following expansion
\begin{equation}
\label{eq:func}
f(x) \,=\, \sum_{n = 0}^\infty\,a_n\,H_n^{(3)}(x,-|y|)\;,
\end{equation}
and we will show that the use of the operational tools developed in this paper provides, in a fairly natural 
way, the coefficients $a_n$,  and we will display the orthogonal properties of this family of polynomials. 
On account of the definition of the higher orders Hermite polynomials (see eq. \eqref{eq:hneg}), 
eq. \eqref{eq:func} yields:
\begin{equation}
\mathrm{e}^{|y| \partial_x^3}\,f(x) \,=\, \sum_{n = 0}^\infty\,a_n\,x^n\;.
\end{equation}
The use of the identity \eqref{eq:fexp} allows to specify the action of  the exponential operator on the 
function $f(x)$ as:
\begin{equation}
\frac{1}{\sqrt[3]{3 |y|}}\,\int_{- \infty}^{\infty}\, \mathrm{d}\xi\, Ai\left(- \frac{x - \xi}{\sqrt[3]{3 |y|}}\right)\,f(\xi)
\,=\, \sum_{n = 0}^\infty\,a_n\,x^n\;,
\end{equation}
and the insertion of the explicit expression of the Airy function (see eq. \eqref{eq:airy}) yields: 
\begin{equation}
\frac{1}{\sqrt[3]{3 |y|}}\,\frac{1}{2 \pi}\,\int_{- \infty}^{\infty}\, \mathrm{d}\xi\,
\int_{- \infty}^{\infty}\, \mathrm{d}t\,\exp{\left\{\frac{\imath}{3}\,t^3\,+\,\imath\,
\frac{\xi - x}{\sqrt[3]{3 |y|}}\,t\right\}}\,f(\xi) \,=\, \sum_{n = 0}^\infty\,a_n\,x^n\;.
\end{equation}
The expansion of the $x$-dependent part of the exponential on the lhs of the previous equation 
yields:
\begin{equation}
\label{eq:acoe}
a_n \,=\,\frac{(-)^n}{2 \pi n!}\,\frac{1}{\sqrt[3]{3 |y|}}\,\int_{- \infty}^{\infty}\, \mathrm{d}t\,
f(\xi)\,\partial_\xi^n\,Ai\left(\frac{\xi}{\sqrt[3]{3 |y|}}\right)\;,
\end{equation}
which holds only if the integral on the rhs of this equation converges.  

In these concluding remarks we have touched on several problems, which are worth to be explored 
thoroughly. The link between the Watson and the $A^{(2p + 1)}(x)$ functions, the possibility of obtaining 
more general Airy-type transforms and their relevance to Bell-type polynomials, and a more rigorous 
formulation of the orthogonality properties of the Airy polynomials will be the topic of a forthcoming 
investigation.

\end{document}